\documentclass[
preprint,
 amsmath,amssymb,
 aps,
prb,
]{revtex4-1}

\usepackage{graphicx}
\usepackage{dcolumn}
\usepackage{bm}
\usepackage{hyperref}
\usepackage[mathlines]{lineno}
\usepackage{commath}
\usepackage{physics}
\usepackage{color}
\usepackage{mathtools}

\begin{document}

\preprint{APS/123-QED}

\title{Large band splitting with tunable spin polarization in two-dimensional ferroelectric Ga$XY$ ($X$= Se, Te; $Y$= Cl, Br, I) family}

\author{Moh. Adhib Ulil Absor}
\email{adib@ugm.ac.id} 
\affiliation{Department of Physics, Universitas Gadjah Mada, Sekip Utara, BLS 21 Yogyakarta Indonesia.}%

\author{Fumiyuki Ishii}%
\affiliation{Nanomaterials Research Institute, Kanazawa University, 920-1192, Kanazawa, Japan.}%

\date{\today}

\begin{abstract}
It has been generally accepted that the spin-orbit coupling effect in noncentrosymmetric materials leads to the band splitting and non-trivial spin polarization in the momentum space. However, in some cases, zero net spin polarization in the split bands may occurs, dubbed as the band splitting with vanishing spin polarization (BSVSP) effect, protected by non-pseudo-polar point group symmetry of the wave vector in the first Brillouin zone [Liu et. al., Nat. Commun. \textbf{10}, 5144 (2019)]. In this paper, by using first-principles calculations, we show that the BSVSP effect emerges in two-dimensional (2D) nonsymmorphic Ga$XY$ ($X$= Se, Te; $Y$= Cl, Br, I) family, a new class of 2D materials having in-plane ferroelectricity. Taking the GaTeCl monolayer as a representative example, we observe the BSVSP effect in the split bands along the $X-M$ line located in the proximity of the conduction band minimum. By using $\vec{k}\cdot\vec{p}$ Hamiltonian derived based on the symmetry analysis, we clarify that such effect is originated from the cancellation of the local spin polarization, enforced by non-pseudo-polar $C_{2v}$ point group symmetry of the wave vector along the $X-M$ line. Importantly, we find that the spin polarization can be effectively induced by applying an external out-of-plane electric field, indicating that an electrically tunable spin polarization for spintronic applications is plausible. 
\end{abstract}

\pacs{Valid PACS appear here}
\maketitle

\section{INTRODUCTION}

The next generation of spintronics relies on the new pathway for manipulating electron's spin without additional external magnetic field, which is achievable by utilizing the effect of spin-orbit coupling (SOC)\cite{Manchon}. In noncentrosymmetric crystalline systems, the SOC leads to an effective magnetic field, $\vec{B}\propto [\nabla V(\vec{r})\times \vec{p}]$, where $V(\vec{r})$ and $\vec{p}$ denote the crystal potential and electron momentum, respectively, that induces the band splitting and non-trivial spin polarization in the momentum space as usually referred as the Rashba \cite{Rashba} and Dresselhaus\cite{Dresselhaus} effects. While the Dresselhauss effect occurs on a system hold bulk inversion asymmetry such as bulk zincblende\cite{Dresselhaus} and wurtzite semiconductors\cite{Wang_Dress}, the Rashba effect has been widely observed on a system having structural inversion asymmetry as previously reported on semiconductor quantum well \cite{Nitta,Caviglia}, surface heavy metal\cite{Koroteev,LaShell}, and several two-dimensional (2D) layered compounds\cite{Zhuang,Popovi,Absor_R,Affandi,Absor_Pol}. Interestingly, it is possible to manipulate the Rashba spin polarization by using an external electric field, offering an opportunity for the realization of spintronic devices such as spin-field effect transistors \cite{Datta,Chuang}.  

From a fundamental point of view, the SOC is a relativistic effect, which strongly depends on the particular atomic-orbital character\cite{Herman}, thereby predominantly sensitives to the local individual atomic sites (called as local real space sectors) in the crystal. Therefore, both the Rashba and Dresselhaus effects can also arise from the local point-group inversion asymmetry of the local real space sectors\cite{Liu2013,Zhang2014,Yuan2019}. In contrast to the conventional Rashba (Dresselhauss) effect, the local Rashba (Dresselhauss) effect is induced by the local dipole fields (site inversion asymmetry), leading to the local spin polarization. Therefore, the superposition of such polarization leads to the total crystalline spin polarization. In centrosymmetric systems, the global inversion symmetry arises but the local real space sector is an inversion asymmetric. As a result, the compensated spin polarization with opposite orientation is degenerate in energy, but is spatially locked to different local sectors of the unit cell called as inversion partners, leading to a trivial (empty) spin polarization of the entire crystal. Such a concept is known as hidden spin polarization effect\cite{Liu2013,Zhang2014,Yuan2019}, as recently observed in various centrosymmetric layered compounds\cite{Yuan2019,Zhang2014,ifmmode,Huang,Yao2017,Liu_Hidde}. More recently, the hidden spin polarization effect protected by nonsymmorphic symmetry in centrosymmetric systems has been reported \cite{Zhang_2020}. 

Analogous to the hidden spin polarization effect in the centrosymmetric systems\cite{Zhang2014,Yuan2019}, a phenomenon dubbed as the band splitting with vanishing spin polarization (BSVSP), i.e., band splitting induced by the global inversion symmetry breaking but with zero net spin polarization, has recently been predicted\cite{Liu2019}. Such a phenomenon, which is occured in noncentrosymmetric system having both the symmorphic and nonsymmorphic symmetries, is strongly different from the conventional Rashba and Dresselhaus effects, where the vanishing spin polarization is protected by non-pseudo-polar symmetry of the little point group. Compared with the conventional Rashba and Dresselhaus effects, the BSVSP effect may have advantages for electrically tunable spintronic devices since the spin polarization can be easily induced by applying an external electric filed\cite{Liu2019}. Therefore, finding novel materials supporting the BSVSP effect for spintronics is very important.
 
In this paper, by performing first-principles density-functional theory calculations, we predict the emergence of the BSVSP effect in 2D nonsymmorphic Ga$XY$ ($X$= Se, Te; $Y$= Cl, Br, I) family, a new class of 2D materials having in-plane ferroelectricity. By using the GaTeCl monolayer (ML) as a representative example, we find that the BSVSP effect is observed in the split bands along the $X-M$ line, which is located in the proximity of the conduction band minimum. By using $\vec{k}\cdot\vec{p}$ Hamiltonian obtained from the symmetry analysis, we confirm that such effect is due to the cancellation of the local spin polarization, suppressed by non-pseudo-polar $C_{2v}$ point group symmetry of the wave vector along the $X-M$ line. Interestingly, we find that significant spin polarization can be induced when an external out-of-plane electric field is applied, indicating that an electrically controllable spin polarization for spintronic applications is plausible. Finally, a possible application of the present system for spintronics will be discussed.

\section{Computational details}

Our first-principles DFT calculations are performed using the generalized gradient approximation (GGA) \cite {Perdew} implemented in the OpenMX code \cite{Openmx}. Here, we adopted norm-conserving pseudo potentials \cite {Troullier} with an energy cutoff of 350 Ry for charge density. The $12\times10\times1$ $k$-point mesh was used. The wave functions were expanded by linear combination of multiple pseudo atomic orbitals generated using a confinement scheme \cite{Ozaki,Ozakikino}, where two $s$-, two $p$-, two $d$-character numerical pseudo atomic orbitals were used. The SOC interaction was included self consistently in all calculations by using $j$-dependent pseudo potentials \citep{Theurich}. 

We deduced the spin vector component ($S_{x}$, $S_{y}$, $S_{z}$) of the spin polarization in the reciprocal lattice vector $\vec{k}$ from the spin density matrix\cite{Kotaka_2013}. The spin density matrix, $P_{\sigma \sigma^{'}}(\vec{k},\mu)$, are calculated using the following relation,  
\begin{equation}
\begin{aligned}
\label{1}
P_{\sigma \sigma^{'}}(\vec{k},\mu)=\int \Psi^{\sigma}_{\mu}(\vec{r},\vec{k})\Psi^{\sigma^{'}}_{\mu}(\vec{r},\vec{k}) d\vec{r}\\
                                  = \sum_{n}\sum_{i,j}[c^{*}_{\sigma\mu i}c_{\sigma^{'}\mu j}S_{i,j}]e^{\vec{R}_{n}\cdot\vec{k}},\\
\end{aligned}
\end{equation}
where $S_{ij}$ is the overlap integral of the $i$-th and $j$-th localized orbitals, $c_{\sigma\mu i(j)}$ is expansion coefficient, $\sigma$ ($\sigma^{'}$) is the spin index ($\uparrow$ or $\downarrow$), $\mu$ is the band index, and $\vec{R}_{n}$ is the $n$-th lattice vector. Here, $\Psi^{\sigma}_{\mu}(\vec{r},\vec{k})$ is the spinor Bloch wave function, which is obtained from the OpenMX calculations after self-consistent is achieved.

\begin{figure}
	\centering		
	\includegraphics[width=0.85\textwidth]{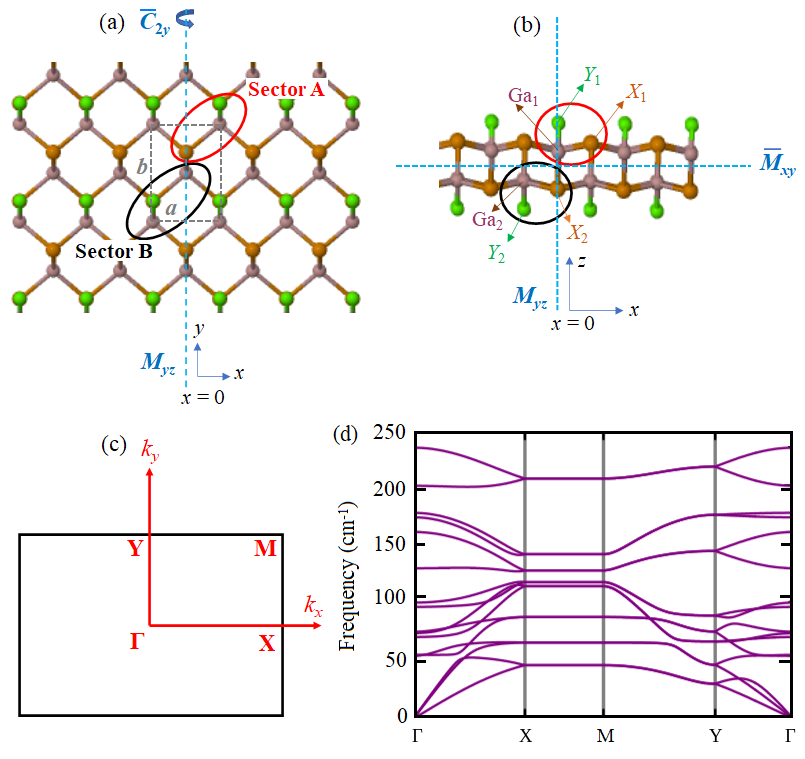}
	\caption{ (a)-(b) Atomic structure of the Ga$XY$ ML corresponding to it symmetry operations viewed in the $x-y$ and $x-z$ planes, respectively. The unit cell of the crystal is indicated by the black dashed lines and characterized by $a$ and $b$ lattice parameters in the $x$ and $y$ directions. The red and black circles show the local sectors $A$ and $B$, respectively. The Ga$_{1}$, $X_{1}$, and $Y_{1}$ indicate the Ga, $X$, and $Y$ atoms located on the sector $A$,  while the Ga$_{2}$, $X_{2}$, and $Y_{2}$ represent the Ga, $X$, and $Y$ atoms located on the sector $B$. (c) First Brillouin zone of the Ga$XY$ ML crystal characterized by the high symmetry $\vec{k}$ points ($\Gamma$, $Y$, $M$, and $X$). (d) Phonon spectrum of the GaTeCl ML as a representative of the Ga$XY$ ML compounds calculated in the FBZ.}
	\label{figure:Figure1}
\end{figure}

We used a periodic slab to model the Ga$XY$ ML, where a sufficiently large vacuum layer (20 \AA) is applied in order to avoid interaction between adjacent layers. We used the axes system where the layer is chosen to sit on the $x-y$ plane [Fig. 1(a)-(b)]. During the structural relaxation, the energy convergence criterion was set to $10^{-9}$ eV. The lattice and positions of the atoms were optimized until the Hellmann-Feynman force components acting on each atom was less than 1 meV/\AA. The phonon spectrum was obtained by using ALAMODE code\cite{Tadano} based on the force constants obtained from the OpenMX code calculations. We used the modern theory of polarization based on the Berry phase method\cite{King-Smith} implemented in the OpenMX code to calculate the spontaneous electric polarization.  

\section{Results and Discussion}

\subsection{Structural symmetry and stability}

First, we analyze the structural symmetry and stability of the Ga$XY$ ML compounds. As shown in Figs. 1(a)-(b) that the crystal structure of the Ga$XY$ ML compounds is noncentrosymmetric having black-phosphorus type structure, where its symmetry is isomorphic to the nonsymmorphic $Pmn2_{1}$ space group\cite{Wu,Kniep,ZhouNEGF,ZhangGaTeCl}. The first Brillouin zone (FBZ) corresponding to this structure is shown in Fig. 1(c). The $Pmn2_{1}$ symmetry in the Ga$XY$ ML is generated by the following symmetry operations [see Fig. 1(a)-(b)]: (i) identity operation $E$, (ii) the glide reflection $\bar{M}_{xy}$ which consists of reflection about $z=0$ plane followed by $a/2$ translation along the $x$ axis and $b/2$ translation along the $y$ axis:
\begin{equation}
\label{2}
\bar{M}_{xy}: (x,y,z)\rightarrow (x+\frac{a}{2}, y+ \frac{b}{2}, -z),
\end{equation}
(iii) the twofold screw rotation $\bar{C}_{2y}$ which consists of $\pi/2$ rotation around $y=b/2$ line followed by $a/2$ translation along the $x$ axis:
\begin{equation}
\label{3}
\bar{C}_{2y}: (x,y,z)\rightarrow (-x+\frac{a}{2}, y+ \frac{b}{2}, -z),
\end{equation}
and (iv) the mirror reflection $M_{yz}$, which is reflection around the $x=0$ plane:
\begin{equation}
\label{4}
M_{yz}: (x,y,z)\rightarrow (-x, y, z).
\end{equation}
Here, $a$ and $b$ is the lattice parameters of the crystal as indicated in Fig. 1(a). 

It is important to note here that there is an intersection site in the Ga$XY$ ML located around the $x=0$ plane, where only the $M_{yz}$ mirror symmetry operation preserves [Fig. 1(a)]. This intersection site, which belongs to $C_{s}$ point group, divides the real space into two sectors, namely $A$ and $B$ sectors. The sector $A$ is consisted of the Ga$_{1}$, $X_{1}$, and $Y_{1}$ atoms, while the sector $B$ is filled by Ga$_{2}$, $X_{2}$, and $Y_{2}$ atoms [Fig. 1(b)]. Although the global atomic site in the Ga$XY$ ML belongs to $C_{2v}$ point group symmetry, which is non-pseudo polar, the local atomic site in each sector reduces to $C_{s}$ point group, which is pseudo polar.

The nonsymmorphic $Pmn2_{1}$ space group symmetry in the Ga$XY$ ML plays an important role for generating the in-plane ferroelectricity\cite{ZhangGaTeCl,ZhouNEGF}. Here, the orientation of the ferroelectric polarization is enforced by the $C_{2v}$ point group related to the $Pmn2_{1}$ space group, similar to that observed on various group IV monochalcogenide monolayers\cite{Fei,Kaloni}. Since the $C_{2v}$ point group contains mirror $xz$ and $yz$ planes, this implies that the net ferroelectric polarization vanishes along the $x$- and $z$-directions, while it is substantial along the $y$-direction. This in-plane ferroelectric polarization is originated from the polar displacements between Ga and $X$ atoms along the $y$-direction. In addition, the existence of the Ga-$Y$ bond in the Ga$XY$ ML contributes to extra dipole moments in the $y$-direction, leading to the large magnitude of the in-plane ferroelectric polarization. 

In this work, we will focus on the GaTeCl ML as a representative example of the Ga$XY$ ML family since the layered GaTeCl bulk material has been experimentally synthesized\cite{Kniep}. Here, the calculated-optimized lattice parameters, $a$ and $b$, are 4.17 \AA\ and 5.93 \AA\, respectively, which is in a good agreement with previous calculation\cite{ZhangGaTeCl,ZhouNEGF}. Our Berry phase calculation found that the calculated in-plane ferroelectric polarization is 597 pC/m, which is consistent well with previous calculation \cite{ZhangGaTeCl,ZhouNEGF}, but is larger than that observed on various 2D in-plane ferroelectric materials \cite{Fei,Ai}. To confirm the structural stability of the GaTeCl ML, we show in Fig. 1(d) the calculated phonon dispersion bands. We can see clearly that there is no imaginary frequencies found in the phonon dispersion bands, indicating that the optimized GaTeCl ML is a dynamically stable.    

\begin{figure}
	\centering		
	\includegraphics[width=1.0\textwidth]{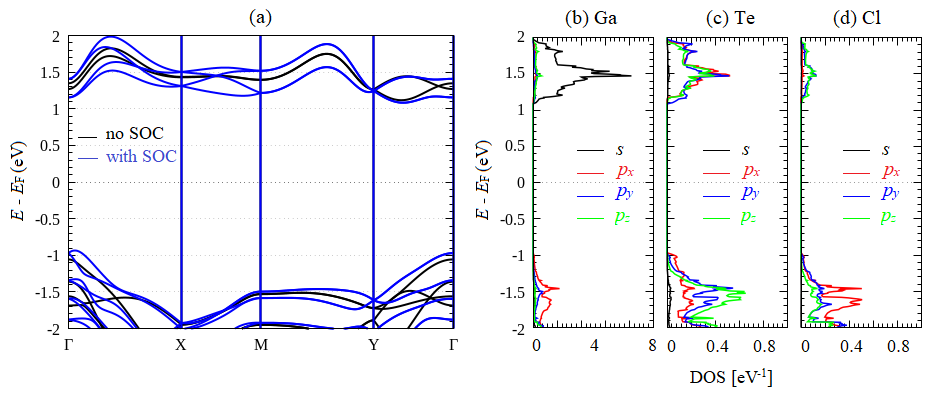}
	\caption{ (a) Electronic band structures of the GaTeCl ML calculated with (blue lines) and without (black lines) including the SOC. Partial density of states projected to the atomic orbital for (b) Ga, (c) Te, and (d) Cl atoms. The black, red, blue, and green lines indicate the $s$, $p_{x}$, $p_{y}$, and $p_{z}$ orbitals, respectively.}
	\label{figure:Figure2}
\end{figure}

\subsection{Band splitting and spin polarization}

Figure 2 shows the electronic band structure of the GaTeCl ML along the selected $\vec{k}$ paths in the FBZ corresponding to the density of states (DOS) projected to the atomic orbitals. One finds that the material is an indirect band-gap semiconductor with the valence band maximum (VBM) is located at the $\Gamma$ point and the conduction band minimum (CBM) is located at the $\vec{k}$ point along the $\Gamma-Y$ line [Fig. 2(a)]. Without including the SOC, the calculated band-gap is 2.17 eV under GGA level, which is in a good agreement with previous calculations\cite{ZhangGaTeCl,ZhouNEGF}. Since there is no magnetic ordering found in the GaTeCl ML, so the time reversal symmetry (TRS) is also preserved. Our calculated DOS projected to the atomic orbitals confirmed that the VBM is mostly dominated by contribution of the Te-$p$ orbital with a small admixture of the Ga-$p$ and Cl-$p$ orbitals, while the CBM is mainly originated from the Ga-$s$ orbital with small contribution of Ga-$p$, Te-$p$ and Cl-$p$ orbitals [Figs. 2(b)-(e)]. 

\begin{figure}
	\centering		
	\includegraphics[width=0.55 \textwidth]{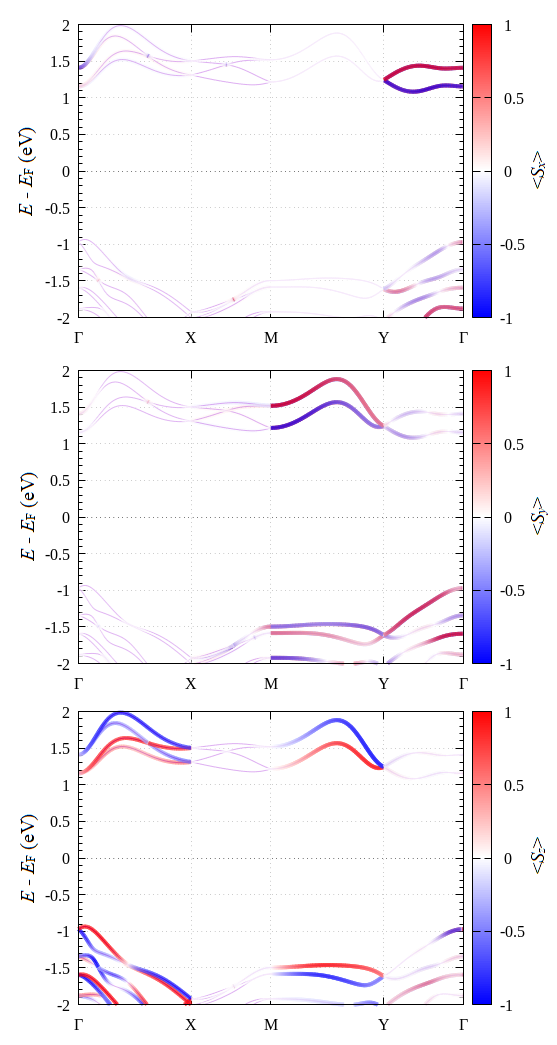}
	\caption{Spin polarization projected to the spin-split bands calculated for $\vec{k}$ along $\Gamma-X-M-Y-\Gamma$ symmetry line. Color bars represent expectation values of spin component $\left\langle S_{x}\right\rangle$, $\left\langle S_{y}\right\rangle$, and $\left\langle S_{z}\right\rangle$. }
	\label{figure:Figure3}
\end{figure}

Turning the SOC, however, slightly reduces the band gap to about 2.05 eV and strongly modifies the electronic band structures of the GaTeCl ML. In comparison with the band structures calculated without SOC [Fig. 2(a)], one can see that a sizable band splitting produced by the SOC is observed due to the inversion symmetry breaking, except for the time-reversal-invariant $\vec{k}$ points. This splitting is particularly visible along the the $\Gamma-X$ and $X-M$ symmetry lines located in the proximity of the CBM. However, due to the protection of the glide mirror symmetry $\bar{M}_{xy}$, the band crossing appears along both the $\Gamma-X$ and $X-M$ symmetry lines, forming a hourglass-shaped band dispersions similar to that observed on bulk BiInO$_{3}$ \cite{Tao2018} and monolayer GaTeI\cite{Wu}. Along the $\Gamma-X$ line, we find that the split bands are fully spin-polarized oriented in the out-of-plane ($z$) direction, while all components of the spin polarization vanish in the split bands along the $X-M$ line [Fig. 3]. The vanishing spin polarization in the split bands along the $X-M$ line indicates that the BSVSP effect is achieved, which is similar to that observed on the SnTe ML\cite{Liu2019}.

The fully out-of-plane spin-polarized bands along the $\Gamma-X$ line can be explained by using unidirectional Rashba effect induced by the in-plane ferroelectricity\cite{Absor_In,Lee,Ai}. The existence of the in-plane ferroelectric polarization along the $y$ direction naturally develops an in-plane electric field, $\vec{E}=E_{y}\hat{y}$, and produces a unidirectional Rashba SOC described by the following Hamiltonian\cite{Absor_In,Lee,Ai}, $H_{\Gamma-X}=\alpha_{R} k_{x}\sigma_{z}$, where $\alpha_{R}$ is the Rashba parameter that is proportional to the magnitude of the in-plane electric field $E_{y}$, and $\sigma_{z}$ is the $z$ component of the Pauli matrices. We noted here that the similar form of $H_{\Gamma-X}$ can also be deduced by considering the little group of the wave vector $\vec{k}$ at the $\Gamma$ point belonging to the $C_{2v}$ point group similar to that reported on various 2D ferroelectric materials such as WO$_{2}$Cl$_{2}$\cite{Ai}, SnTe \cite{Absor_In,Lee}, and SnSe\cite{ANSHORY} MLs. Therefore, the spin orientation along the $\Gamma-X$ ($k_{x}$) line is completely locked in the out-of-plane direction, forming a fully unidirectional out-of-plane spin polarization [Fig. 3]. This spin polarization is expected to maintain the persistent spin helix state \cite{Bernevig,Schliemann} along the $\Gamma-X$ line, which suppresses the DP mechanism of the spin relaxation\cite{Dyakonov} and induces an extremely long spin lifetime\cite{Altmann}. 

The vanishing spin polarization as a manifestation of the BSVSP effect in the split bands along the $X-M$ symmetry line, on the other hand, cannot be explained in term of the unidirectional Rashba effect. Recently, based on the group theory analysis, the emergence of the BSVSP effect has been proposed for both symmorphic and non-symmorphic systems satisfying the following two-simultaneous conditions\cite{Liu2019} : (i) the little space group associated with the wave vector $\vec{k}$ possess 1D double-value irreducible representation (IR); (ii) the little point group associated with $\vec{k}$ should be a non-pseudo polar point group (detailed analysis based on the point group theory can be found in the supplementary of Ref. 2). In our system, the crystal symmetry is isomorphic to the $Pmn2_{1}$ space group whose corresponds to the point group $C_{2v}$\cite{Wu}. Along the $X-M$ line, the little space group of $\vec{k}$ belongs to $C_{2v}$ point group. Therefore, we find that there exists 1D double-value IR of the little space group along the $X-M$ line. At the same time, the little point group of $\vec{k}$ along the $X-M$ line also belongs to $C_{2v}$ point group, which is non-pseudo polar. All these facts confirmed that the split bands along the $X-M$ line sustains the BSVSP effect [see Fig. 3].

\begin{figure}
	\centering		
	\includegraphics[width=1.0\textwidth]{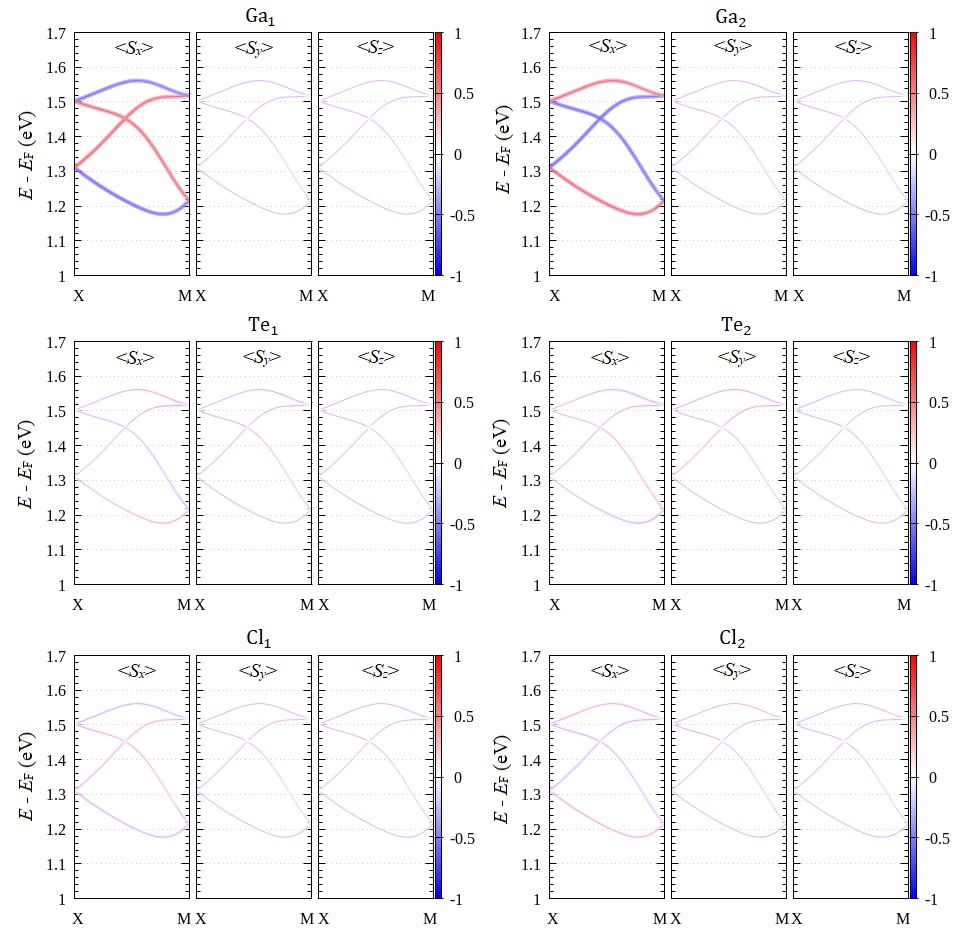}
	\caption{ Atomic decomposition of the spin polarization projected to the spin-split bands along the $X-M$ line. The Upper, middle, and lower panels show the decomposition of the spin polarization on the Ga, Te, and Cl atomic pairs in the unit cell, respectively. Color bars represent expectation values of spin component $\left\langle S_{x}\right\rangle$, $\left\langle S_{y}\right\rangle$, and $\left\langle S_{z}\right\rangle$. }
	\label{figure:Figure4}
\end{figure}

To clarify the origin of the BSVSP effect along the $X-M$ line, we evaluate the atomic contribution on the spin polarization in the spin-split bands. Here, we focused on the spin-split bands along the $X-M$ line at the CBM due to the large spin splitting [Fig. 2(a)]. Since there are six atoms in the unit cell of the GaTeCl ML, i.e., the Ga$_{1}$, Te$_{1}$, and Te$_{1}$ atoms in the sector $A$; and the Ga$_{2}$, Te$_{2}$, and Cl$_{2}$ atoms in the sector $B$ [Fig. 1(b)], we then decompose the spin polarization of the selected bands into these atoms. As shown in Fig. 4, we find that the spin polarizations are dominated by the $S_{x}$ component of spin, which is mainly originated from the contribution of the Ga atoms. These spin polarizations are opposite in the $x$ direction when projected into the Ga atom at the different sectors, indicating that the spin polarizations are locally emerged. Such local and opposite spin polarizations found in the bands along the $X-M$ line are analogous to the hidden spin-polarization effect\cite{Zhang2014,Yuan2019} observed on the centrosymmetric layer compounds\cite{Yuan2019,Zhang2014,ifmmode,Huang,Yao2017,Liu_Hidde}. Since the orientations of the local spin polarization are opposite in the $x$ direction, they should cancel each other so that the net spin polarizations is zero, thus maintaining the BSVSP effect in agreement with total spin polarization shown in Fig. 3. 

We noted here that the vanishing spin polarization can be understood by considering the site symmetry of the atoms in each sector in the unit cell of the GaTeCl ML. Due to the existence of the $M_{yz}$ mirror symmetry operation [Fig. 1(a)-(b)], only the $x$-component of spin polarizations preserves, while the $y$- and $z$-components of the spin polarizations are zero. However, these spin polarizations appear locally on each sector, which is suppressed by the pesudo polar $C_{s}$ point group of the intersection site symmetry. Since the atomic sites in each sector are also characterized  by the pseudo polar $C_{s}$ point group symmetry, the $x$-component of the spin polarization also maintains locally on each atom. Considering the fact that the spin-split bands along the $X-M$ line at the CBM are mainly originated from the Ga-$s$ orbital [Fig. 2(b)-(d)], their local spin polarizations are dominated by the contribution of the Ga atoms rather than Te and Cl atoms. However, these local spin polarizations have opposite signs when projected to the Ga atom in the different sectors [Fig. 4], thus resulting in vanishing of the net spin polarization as shown in Fig. 3.

To further demonstrate the microscopic origin of the BSVSP effect along the $X-M$ line at the CBM, we construct $\vec{k}\cdot\vec{p}$ effective Hamiltonian describing the band structure around the high symmetry $X$ point. Here, the $\vec{k}\cdot\vec{p}$ Hamiltonian can be derived based on the theory of invariant\cite{Luttinger,Winkler}. For the particular high symmetry point in the first Brillouin zone, the little group of the wave vector $\vec{k}$ is characterized by a point group $G$, where the matrix representation for the chosen basis functions is given by $\left\{D(g):g\in G\right\}$, where $g$ is the symmetry operations in the point group. The derived Hamiltonian should satisfy the following invariant condition\cite{Winkler}:
\begin{equation}
\label{5}
H(\vec{k})=D(g)H(g^{-1}\vec{k})D^{-1}(g), \ \ \forall g\in G.
\end{equation}

\begin{table}[ht!]
\caption{Transformation role for wave vector ($k$), spin ($\sigma$), and sublattice ($\tau$) Pauli matrices under $C_{2v}$ point group symmetry operations for the $X$ point in the first Brillouin zone. $K$ denotes complex conjugation.} 
\centering 
\begin{tabular}{ccc ccc ccc  ccc} 
\hline\hline 
 Symmetry operation            &&& ($k_{x}$, $k_{y}$)   &&&   ($\sigma_{x}$, $\sigma_{y}$, $\sigma_{z}$)  &&& ($\tau_{x}$, $\tau_{y}$, $\tau_{z}$) \\ 
\hline 
$T=i\sigma_{y}\tau_{z}K$       &&& ($-k_{x}$, $-k_{y}$)  &&&  ($-\sigma_{x}$, $-\sigma_{y}$, $-\sigma_{z}$)    &&&   ($-\tau_{x}$, $\tau_{y}$, $\tau_{z}$)   \\  
$C_{2y}=i\sigma_{y}\tau_{x}$   &&& ($-k_{x}$, $k_{y}$)   &&&   ($-\sigma_{x}$, $\sigma_{y}$, $-\sigma_{z}$)  &&&   ($\tau_{x}$, $-\tau_{y}$, $-\tau_{z}$)   \\  
$M_{yz}=\sigma_{x}\tau_{x}$    &&& ($-k_{x}$, $k_{y}$)   &&&    ($\sigma_{x}$, $-\sigma_{y}$, $-\sigma_{z}$)   &&&    ($\tau_{x}$, $-\tau_{y}$, $-\tau_{z}$) \\ 
$M_{xy}=i \sigma_{z}$          &&& ($k_{x}$, $k_{y}$)    &&&    ($-\sigma_{x}$, $-\sigma_{y}$, $\sigma_{z}$)   &&&    ($\tau_{x}$, $\tau_{y}$, $\tau_{z}$)   \\ 
\hline\hline 
\end{tabular}
\label{table:Table 1} 
\end{table}

At the $X$ point, the little group of the wave vector $\vec{k}$ belongs to the $C_{2v}$ point group. Therefore, the wave vector $\vec{k}$ and spin vector $\vec{\sigma}$ can be transformed according to the symmetry operations in this point group. Taking into account the spin and sub lattice degree of freedom in the $X$ point, the corresponding transformations are given in Table I. Collecting all terms which are invariant under these symmetry operations, we obtain the following effective Hamiltonian:
\begin{equation}
\label{6}
H_{X} = E_{0}+\lambda\tau_{z}\sigma_{z}+ \alpha k_{y}\tau_{y}\sigma_{x}+\beta k_{x}\tau_{y}\sigma_{y}+\gamma k_{x} \tau_{0}\sigma_{z}+\delta k_{x}\tau_{z}\sigma_{0}, 
\end{equation}
where $E_{0}=\frac{\hbar^{2}k^{2}_{x}}{2m_{x}}+ \frac{\hbar^{2}k^{2}_{y}}{2m_{y}}$ is the free electron contribution with $m_{x,y}$ being the effective mass along the $k_{x}$ and $k_{y}$ directions. Here, $\lambda$, $\alpha$, $\beta$, $\gamma$, and $\delta$ are independent parameters, $\sigma_{0}$ and $\tau_{0}$ are the $2\times2$ identity matrices. Along the $X-M$ line, we have $k_{x}=0$, so the Hamiltonian of Eq. (\ref{6}) can be written as
\begin{equation}
\label{7}
H_{X-M} = E_{0_{y}}+\lambda\tau_{z}\sigma_{z}+\alpha k_{y}\tau_{y}\sigma_{x}, 
\end{equation} 
where $E_{0_{y}}=\frac{\hbar^{2}k^{2}_{y}}{2m_{y}}$. At the $X$ point, the $\lambda\tau_{z}\sigma_{z}$ term in Eq. (\ref{7}) splits the states into two doublets ($\Psi_{1}$,$\Psi_{2}$) separated by $\Delta = 2 \lambda$. Solving eigenvalue problem involving the Hamiltonian of Eq. (\ref{7}) leads to the solution for each doublets as follows:
\begin{equation}
\begin{aligned}
\label{8}
\Psi^{\pm}_{1}(\vec{k})= \frac{e^{\vec{k}\cdot\vec{r}}}{\sqrt{2}}\left(\begin{array}{c}
\pm 1\\
 1 \\
\end{array} \right),  \ \ E^{\pm}_{\Psi_{1}}=E_{0_{y}} + \lambda \pm \alpha k_{y}\\
\Psi^{\pm}_{2}(\vec{k})= \frac{e^{\vec{k}\cdot\vec{r}}}{\sqrt{2}}\left(\begin{array}{c}
\mp 1\\
 1 \\
\end{array} \right),  \  \ E^{\pm}_{\Psi_{2}}=E_{0_{y}} - \lambda \pm \alpha k_{y}.\\
\end{aligned}
\end{equation}
By fitting the band dispersion of Eq. (\ref{8}) to the DFT energy band along the $X-M$ in the CBM, we obtain the following parameters: $\lambda=0.09$ eV and $\alpha=2.27$ eV\AA. Importantly, the calculated $\alpha$, which represents the spin-orbit strength parameter, is much larger than that observed on various 2D materials\cite{Zhuang,Popovi,Absor_R,Affandi,Absor_Pol}, which is sufficient to support the room temperature spintronics functionality.

The BSVSP effect can be further confirmed by calculating the spin polarization projected to the atoms in each sector in the unit cell. As previously mentioned that there are two local sectors in the unit cell of the GaTeCl ML, where the intersection site is characterized by only the $M_{yz}$ mirror symmetry operation [Fig. 1(a)]. Therefore, we define the projected wave function onto each sector as $\Psi^{A(B)}(\vec{k})=P^{A(B)}\Psi(\vec{k})$, where $P^{A(B)}$ is the projection operator onto $A(B)$ sectors and $\Psi(\vec{k})$ is the global wave function given in Eq. (\ref{8}). The expectation value of spin operator projected to the sectors $A(B)$ can be written as
\begin{equation}
\label{8a}
\left\langle \vec{S} \right\rangle^{A(B)} =\frac{\hbar}{2} \left\langle \Psi^{A(B)}(\vec{k})|\vec{\sigma}| \Psi^{A(B)}(\vec{k}) \right\rangle
\end{equation}    
By using the fact that $\Psi(\vec{k})$ is related to the $\Psi(-\vec{k})$ by the mirror symmetry $M_{yz}$ and  time reversal symmetry $T$ operations, we obtain that
\begin{equation}
\begin{split}
\label{8b}
\left\langle \vec{S} \right\rangle^{A} & =\frac{\hbar}{2} \left\langle \Psi^{A}(\vec{k})|\vec{\sigma}| \Psi^{A}(\vec{k}) \right\rangle\\
    & = \frac{\hbar}{2} \left\langle \Psi(\vec{k})|P^{A} \vec{\sigma}P^{A}| \Psi(\vec{k}) \right\rangle\\
    & = \frac{\hbar}{2} \left\langle T \Psi(\vec{k})|M_{yz}^{-1} M_{yz} T P^{A} \vec{\sigma}P^{A} T^{-1} M_{yz}^{-1} M_{yz} | T \Psi(\vec{k}) \right\rangle\\
    & = \frac{\hbar}{2} \left\langle \Psi(-\vec{k})|M_{yz}^{-1} P^{B} \vec{\sigma}P^{B} M_{yz} | \Psi(-\vec{k}) \right\rangle\\
   &  =-\frac{\hbar}{2} \left\langle \Psi(\vec{k})| P^{B} \vec{\sigma}P^{B} | \Psi(\vec{k}) \right\rangle\\
 & =  -\frac{\hbar}{2} \left\langle \Psi^{B}(\vec{k})| \vec{\sigma} | \Psi^{B}(\vec{k}) \right\rangle\\
&  = -\left\langle \vec{S} \right\rangle^{B}.\\
\end{split} 
\end{equation}   
We can clearly see that the expectation value of the spin operator shows an opposite sign between $A$ and $B$ sectors, indicating that the local spin polarizations have opposite orientation. 

Now, we focused on the expectation value of the spin operator projected to the Ga atom since the spin-split bands along the $X-M$ line is mainly dominated by the Ga-$s$ orbitals [Fig. 2(b)]. By evaluating Eq. (\ref{8b}) together with the wave functions given in Eq. (\ref{8}), we find that the expectation value of the spin operator projected to the Ga atoms can be written as
\begin{equation}
\begin{aligned}
\label{9}
\left( \left\langle S_{x}\right\rangle,\left\langle S_{y}\right\rangle, \left\langle S_{z}\right\rangle \right)^{\pm}_{\Psi^{\pm, Ga_{1}}_{1}} = \pm \frac{\hbar}{2}(1,0,0) \\
\left( \left\langle S_{x}\right\rangle,\left\langle S_{y}\right\rangle, \left\langle S_{z}\right\rangle \right)^{\pm}_{\Psi^{\pm, Ga_{2}}_{1}} = \mp \frac{\hbar}{2}(1,0,0)\\
\left( \left\langle S_{x}\right\rangle,\left\langle S_{y}\right\rangle, \left\langle S_{z}\right\rangle \right)^{\pm}_{\Psi^{\pm, Ga_{1}}_{2}} = \mp \frac{\hbar}{2}(1,0,0)\\
\left( \left\langle S_{x}\right\rangle,\left\langle S_{y}\right\rangle, \left\langle S_{z}\right\rangle \right)^{\pm}_{\Psi^{\pm, Ga_{2}}_{2}} = \pm \frac{\hbar}{2}(1,0,0).\\
\end{aligned} 
\end{equation}    
This shows that both Ga$_{1}$ and Ga$_{2}$ atoms contributes to the local spin polarization having opposite direction along the $x$ direction, which is consistent well with our DFT results shown in Fig 4. Therefore, the net spin polarization becomes zero, giving rise to the BSVSP effect that is in agreement with our DFT results presented in Fig. 3.  

\begin{figure}
	\centering		
	\includegraphics[width=1.0\textwidth]{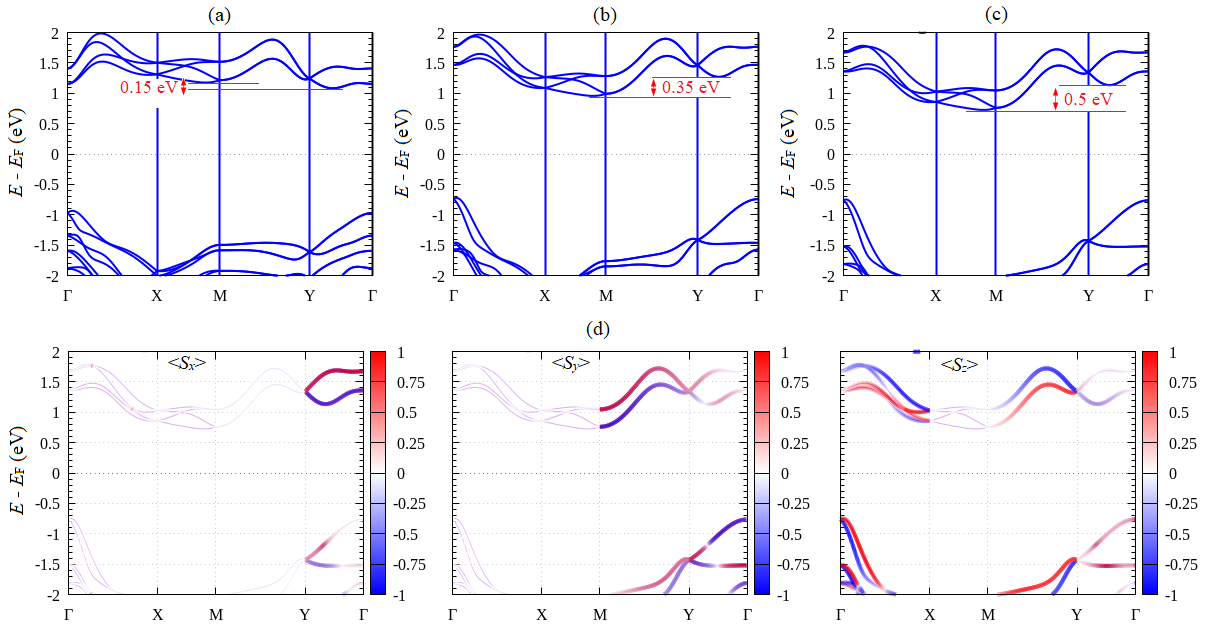}
	\caption{Evolution of the band structures under small compressive uniaxial strain along the $x$ (zigzag) direction: (a) equilibrium, (b) -4\%, and (d) -8\%. Introducing the strain significantly modifies the electronic band structures where the CBM shifts from the $k$ point at the $\Gamma-Y$ line to that at the $X-M$ line as highlighted by the red lines. (d) Spin polarization projected to the spin-split bands along $\Gamma-X-M-Y-\Gamma$ symmetry line for the case of the compressive uniaxial strain of -8\%.  Color bars represent expectation values of the spin components $\left\langle S_{x}\right\rangle$, $\left\langle S_{y}\right\rangle$, and $\left\langle S_{z}\right\rangle$. }
	\label{figure:Figure5}
\end{figure}

We emphasized here that the observed BSVSP effect in the present system is located in the spin-split bands along the $X-M$ line, which is slightly higher in energy (0.15 eV) than the CBM at the $k$ point along the $\Gamma-Y$ line [Fig. 5(a)]. However, one has to note that the CBM can be effectively tuned by a small amount of strain [Fig. 5(b)-(c)], which is realized by choosing appropriate substrates. For example, under compressive strain of -8\% along the $x$ direction (zigzag), the CBM shifts significantly (with shifting energy of 0.5 eV) being located at the $k$ point along the $X-M$ line [see Fig. 5(c)] without losing the BSVSP effect [Fig. 5(d)]. Therefore, it is possible to access the BSVSP effect experimentally by introducing electron doping. Recently, the electron doping techniques in various 2D materials are under rapid development realized by using ion liquid gating technique \cite{Mak2013,Zhang725}, thus application of the electron doping in GaTeCl ML is also plausible. Since the BSVSP effect in the strained GaTeCl ML is observed in the proximity of the CBM, introducing n-type electron doping can move the interesting bands close to the Fermi level, which can be further resolved by using spin-polarized angle-resolved photo-electron spectroscopy. 

Finally, we highlight the main difference of the BSVSP effect found in the present system with that observed on SnTe ML as previously reported by Liu et. al.\cite{Liu2019}. In the SnTe ML, the BSVSP effect is predicted to appear on the spin-split bands located at valley near the $X$ point in both the CBM and VBM \cite{Liu2019}. However, due to the orthorhombic structure, there is another valley, namely $Y$ valley, characterized the electronic band structures of the SnTe ML \cite{Absor_In, Kim2019}. Since the energy band of both the $X$ and $Y$ valleys are very close, the observed BSVSP effect near the $X$ valley may be interfered by the spin-polarized states in the spin-split bands at $Y$ valley. Therefore, it is difficult to exclusively detected the BSVSP effect without mixing with the spin-polarized states. In contrast, the observed BSVSP effect in the present system is located at the non-valley bands in the proximity of the CBM where the mixing with spin-polarized states can be minimized, for example, by introducing the small-compressive uniaxial strains [Fig.5(b)-(c)]. Moreover, compared with the SnTe ML, the observed BSVSP effect in the present system exhibits the larger spin splitting, which is rationalized by the fact that the spin-orbit strength parameter predicted in the present system (2.27 eV\AA) is much larger than that observed in the SnTe ML (1.2 eV\AA )\cite{Absor_In}. Therefore, the present system provides an advantage to higher possibility for realization spintronic devices operating at room temperature.

\subsection{Tunable spin polarization by an external electric filed}

Although the observed BSVSP effect exhibits the large bands splitting along the $X-M$ line, which is beneficial for spintronics, the vanishing spin polarization in these bands may induce the undesired effect of losing the spin information. Therefore, inducing the spin polarization is the important task for realization spintronic devices. Since the BSVSP effect along the $X-M$ line is protected by the non-pseudo-polar $C_{2v}$ point group, reducing this point group symmetry may exhibits the non-zero spin polarization. For this propose, we apply an external out-of-plane electric field $\vec{E}=E_{z}\hat{z}$ perpendicular to the GaTeCl ML thin film, which can be realized by introducing a gate voltage. Introducing the out-of-plane external electric field is expected to break both the glide mirror plane $\bar{M}_{xy}$ and two-fold screw rotation axis $\bar{C}_{2y}$ in the crystal of the GaTeCl ML, implying that the point group symmetry reduces to the pseudo-polar $C_{s}$ point group. Therefore, the non-zero spin-polarization is allowed in the bands along the $X-M$ line. 

Fig. 6(a) shows the calculated spin polarization projected to the bands along the $X-M$ line under the influenced of 1.0 V/\AA\ out-of-plane electric field. By comparing the bands with and without an external electric field [see Fig. 3 and Fig. 6(a)], we find that the band crossing at the $\vec{k}$ along the $X-M$ line breaks due to the external electric field, so that the Hourglass band dispersion splits into two pair of the split bands. In contrast to the system without the external electric field, the split bands along the $X-M$ line exhibit significant spin polarization, which is dominated by the $S_{x}$ component of spin [Fig. 6(a)]. Moreover, by evaluating the spin polarization at certain $\vec{k}$ along the $X-M$ line, one can observe that the magnitude of the spin polarization enhances linearly by increasing the magnitude of the external electric field [Fig. 6(b)], indicating that the spin polarization can be effectively tuned by an external electric field. Interestingly, we find that the orientation of the spin polarizations can be fully reversed by switching the direction of the external electric field. These tunable and reversible spin polarization by the external electric field enable the controlled of the spin-dependent properties, which is the hint for realization spintronic devices. 

The large spin polarizations in the split bands along the $X-M$ line induced by the out-of-plane external electric field can further be understood by considering the $\vec{k}\cdot\vec{p}$ Hamiltonian around the $X$ point. By including the out-of-plane electric field contribution, the total $\vec{k}\cdot\vec{p}$ Hamiltonian can rewritten as
\begin{equation}
\label{10}
H_{X}^{E_{z}} = H_{X}+H_{E_{z}}=H_{X}+\alpha_{E_{z}}\tau_{y}\sigma_{0},
\end{equation}   
where $H_{X}$ is given in Eq. (\ref{6}) and $\alpha_{E_{z}}=E_{z}\left\langle\phi(\vec{k},\vec{r})|z|\phi(\vec{k},\vec{r})\right\rangle$. Here, $E_{z}$ is the magnitude of the out-of-plane external electric field. Along the $X-M$ line, the Hamiltonian of Eq. (\ref{10}) can be simplified to
\begin{equation}
\label{11}
H_{X-M}^{E_{z}} = E_{0_{y}}+\lambda\tau_{z}\sigma_{z}+\alpha k_{y}\tau_{y}\sigma_{x} + \alpha_{E_{z}}\tau_{y}\sigma_{0}. 
\end{equation}
This Hamiltonian leads to the solutions:
\begin{equation}
\begin{aligned}
\label{12}
\Psi^{\pm}_{1}= \frac{e^{i\vec{k}\cdot\vec{r}}}{\sqrt{\left( \frac{(\alpha k_{y})^{2}+\alpha^{2}_{E_{z}}}{\left(\sqrt{(\alpha k_{y})^{2}+\alpha^{2}_{E_{z}}} \pm \lambda \right)^{2}} \right) + 1 }}   
\left(\begin{array}{c}
\pm \frac{\sqrt{(\alpha k_{y})^{2}+\alpha^{2}_{E_{z}}}} {\sqrt{ (\alpha k_{y})^{2} + \alpha^{2}_{E_{z}}}    \pm \lambda } \\
 1 \\
\end{array} \right),  \ \ E^{\pm}_{\Psi_{1}}=E_{0_{y}} + \lambda \pm \sqrt{(\alpha k_{y})^{2}+\alpha^{2}_{E_{z}}}\\
\Psi^{\pm}_{2}= \frac{e^{i\vec{k}\cdot\vec{r}}}{\sqrt{\left( \frac{(\alpha k_{y})^{2}+\alpha^{2}_{E_{z}}}{\left(\sqrt{(\alpha k_{y})^{2}+\alpha^{2}_{E_{z}}} \mp \lambda \right)^{2}} \right) + 1 }}   
\left(\begin{array}{c}
\mp \frac{\sqrt{(\alpha k_{y})^{2}+\alpha^{2}_{E_{z}}}} {\sqrt{ (\alpha k_{y})^{2} + \alpha^{2}_{E_{z}}}    \mp \lambda } \\
 1 \\
\end{array} \right),  \  \ E^{\pm}_{\Psi_{2}}=E_{0_{y}} - \lambda \pm \sqrt{(\alpha k_{y})^{2}+\alpha^{2}_{E_{z}}}.\\
\end{aligned}
\end{equation}
For the case of the external electric field $E_{z}=1.0$ V/\AA\, the parameters $\lambda$, $\alpha$, and $\alpha_{E_{z}}$ can be obtained by fitting the energy dispersion of Eq. (\ref{12}) to the DFT energy band of Fig. 6(a), and we find that $\lambda=0.11$ eV, $\alpha=2.12$ eV\AA\, and $\alpha_{E_{z}}=0.03$ eV. 

\begin{figure}
	\centering
		\includegraphics[width=0.55\textwidth]{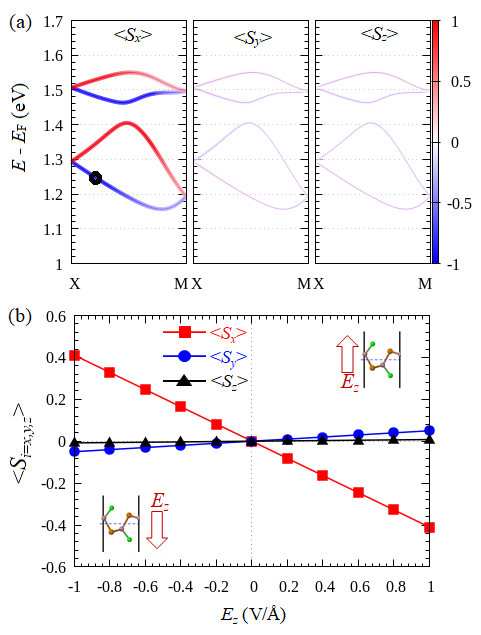}
	\caption{Electric field effect on the spin polarization. (a) The spin polarization projected to the band structures of GaTeCl ML along the $X-M$ line an external out-of-plane electric electric, filed $E_{z}=1.0$ V/\AA\,  is shown. The colours indicate spin polarization components ($S_{x}$, $S_{y}$, $S_{z}$). (b) The calculated spin polarization as a function of out-of-plane electric field. The spin polarizations are calculated at certain the $\vec{k}$ point in the lowest conduction band along the $X-M$ line as indicated in black point in Fig. 6(a).}
	\label{fig:Figure6}
\end{figure}

Furthermore, by using the wave function given on Eq. (\ref{12}), the calculated expectation values of the spin operators can be expressed as
\begin{equation}
\begin{aligned}
\label{13}
\left( \left\langle S_{x}\right\rangle,\left\langle S_{y}\right\rangle, \left\langle S_{z}\right\rangle \right)^{\pm}_{\Psi^{\pm}_{1}} = \pm \frac{\hbar \alpha_{E_{z}} \sqrt{ \frac{(\alpha k_{y})^{2}}{\alpha^{2}_{E_{z}}} +1} }{2   \sqrt{(\alpha k_{y})^{2}+ \alpha^{2}_{E_{z}} }  \pm \lambda }(1,0,0) \\
\left( \left\langle S_{x}\right\rangle,\left\langle S_{y}\right\rangle, \left\langle S_{z}\right\rangle \right)^{\pm}_{\Psi^{\pm}_{2}} = \mp \frac{\hbar \alpha_{E_{z}} \sqrt{ \frac{(\alpha k_{y})^{2}}{\alpha^{2}_{E_{z}}} +1} }{2   \sqrt{(\alpha k_{y})^{2}+ \alpha^{2}_{E_{z}} }  \mp \lambda }(1,0,0). \\
\end{aligned} 
\end{equation} 
It is clearly seen from Eq. (\ref{13}) that the electric field induces the significant spin polarization along the $x$ direction, which is consistent well with our DFT results shown in Fig. 6(a). Moreover, switching the direction of the electric field from $E_{z}$ to $-E_{z}$ revers the sign of $\alpha_{E_{z}}$ to $-\alpha_{E_{z}}$, leading to reversing the spin polarization  in the $x$ direction that is also in agreement with our DFT results shown in Fig. 6(b). 
 
Now, we discuss the possible applications of the electric field-driven spin polarization in GaTeCl ML. Here, we propose a possibility to observe a Hall effect similar to the valley Hall effect previously reported on the 2D transition metal dichalcogenides\cite{Mak1489}. As mentioned previously that introducing an external electric field in the GaTeCl ML produces the large spin polarization in the split bands along the $X-M$ line at the CBM [Fig. 6(a)-(b)]. Accordingly, such spin polarization is expected to occur in the two states located at the $\vec{k}$ and $-\vec{k}$ near the CBM. This implies that the Berry curvatures should be observed with opposite sign. By using polarized optical excitation, it is possible to create imbalance population of the electron in these two states, and hence a charge Hall current can be detected. 

Finally, we discuss another possible application of the  GaTeCl ML in term of the reversible spin polarization effect induced by switching the external electric field. Such reversible spin polarization may be implemented in the magnetic tunnel junctions\cite{Tsymbal_2003}, where at the interface of a magnetic tunnel junction, the Rashba SOC induces a tunneling spin Hall effect and tunneling anomalous Hall effect (AHE)\cite{Vedyayev}. Since the magnitude of the tunneling AHE is linearly depends on the SOC parameter, the tunneling AHE effect is more experimentally accessible for the systems having larger the SOC parameter. In our system, we found that large the spin-orbit parameter ($\alpha=2.12$ eV\AA) is observed when the electric field $E_{z}=1.0$ V/\AA\, indicating that this material is a favorable candidate for detecting the AHE effect experimentally. Since, in our system, the spin polarization can be reversed by switching the direction of the electric field, it is expected that the reversible AHE effect by the electric field is also achieved. Therefore, we conclude that the present system is promising candidate for spintronic applications. 

\section{CONCLUSION}

In summary, we have investigated the emergence of the BSVSP effect in the 2D nonsymmorphic Ga$XY$ ($X$= Se, Te; $Y$= Cl, Br, I) family by performing first-principles density-functional theory calculations. By considering the GaTeCl ML as a representative example, we have found that the BSVSP effect is observed in the split bands along the $X-M$ line in the proximity of the CBM. By deriving the $\vec{k}\cdot\vec{p}$ Hamiltonian obtained from the symmetry analysis, we have confirmed that the BSVSP effect along the $X-M$ line is originated from the cancellation of the local spin polarization, enforced by the non-pseudo-polar $C_{2v}$ point group symmetry of the wave vector $\vec{k}$. We also found that large spin polarization in the split bands along the $X-M$ line can be effectively induced by applying an external out-of-plane electric field, thus offering an electrically controllable spin polarization for spintronic applications. 

The BSVSP effect found in the present study is solely dictated by the non-pseudo-polar $C_{2v}$ point group of the wave vector in the systems having the non-symmorphic $Pnm2_{1}$ space group symmetry. Therefore, we expect that the BSVSP effect discussed here is also shared by other materials having the similar symmetry. These allowed us to implement our analysis provided here to be directly applied to these materials. Recently, there are a number of other 2D materials that are predicted to maintain the similar symmetry of the crystals, which opens a possibility to further to resolve the BSVSP effect in these materials. For example, the better resolved of the BSVSP effect are expected to occur in 2D single-elemental multiferroic materials such as As, Sb, and Bi due to the stronger SOC\cite{Xiao,Pan}. Therefore, we expect that our predictions will stimulate further theoretical and experimental efforts in the exploration of the BSVSP effect in the 2D-based ferroelectric materials, broadening the range of the 2D materials for future spintronic applications.

\begin{acknowledgments}

This work was partly supported by the BPPTNBH Research grant (2020) funded by Faculty of Mathematics and Natural Sciences, Universitas Gadjah Mada, Republic of Indonesia. Part of this research was supported by PD Research Grant (1950/UN1/DITLIT/DIT- LIT/PT/2020) and PDUPT research grant (1950/UN1/DITLIT/DIT- LIT/PT/2020) funded by RISTEK-BRIN, Republic of Indonesia. This work was partly supported by Grants-in-Aid for Scientific Research (Grant No. 16K04875) from JSPS and Grant-in-Aid for Scientific Research on Innovative Areas Discrete Geometric Analysis for Materials Design (Grant No. 18H04481) from MEXT Japan. The computation in this research was performed using the supercomputer facilities at RIIT, Kyushu University, Japan.

\end{acknowledgments}

\bibliography{Reference1}


\end{document}